\documentclass[aps,pra,preprint,showpacs,superscriptaddress]{revtex4}

\usepackage{graphicx}
\usepackage{amsmath}
\usepackage{amssymb}
\input{epsf}
\begin{document}
\providecommand{\bp}{{\bf p}}
\providecommand{\bq}{{\bf q}}
\providecommand{\bk}{{\bf k}}
\providecommand{\bkp}{{\bf k'}}
\providecommand{\xf}{\xi^F}
\providecommand{\xm}{\xi^m}
\providecommand{\w}{\omega}

\title{Generalized Mean Field Approach to a Resonant Bose-Fermi Mixture}

\author{D.C.E.  Bortolotti}
\affiliation{JILA, NIST and Department of Physics, University of Colorado,
Boulder, CO 80309-0440}
\affiliation{LENS and Dipartimento di Fisica, Universit\'a di Firenze,
and INFM, Sesto Fiorentino, Italy}
\author{A.V. Avdeenkov}
\affiliation{Institute of Nuclear Physics, Moscow State University,
Vorob'evy Gory, 119992, Moscow, Russia  }
\author{J.L. Bohn}
\affiliation{JILA, NIST and Department of Physics, University of Colorado,
Boulder, CO 80309-0440}

\date{\today}

\begin{abstract}
We formulate a generalized mean-field theory of a mixture of
fermionic and bosonic atoms, in which the fermion-boson interaction
can be controlled by a Feshbach resonance.  The theory correctly
accounts for molecular binding energies of the molecules in the
two-body limit, in contrast to the most straightforward mean-field
theory.  Using this theory, we discuss the equilibrium properties of
fermionic molecules created from atom pairs in the gas.  We also
address the formation of molecules when the magnetic field is ramped
across the resonance, and present a simple Landau-Zener result for
this process.

\end{abstract}


\maketitle
\section{introduction}

The use of magnetic Feshbach resonances to manipulate the interactions in
ultracold quantum gases has greatly enriched the study of many-body
 physics. Notable examples include
the crossover between BCS (Bardeen-Cooper-Schrieffer \cite{BCS1957}) and BEC
(Bose-Einstein Condensate \cite{bose1924,einstein1924}) superfluidity
in ultracold Fermi gases \cite{regal2003nature_sup,bartenstein2004cmb,
  zwierlein2004cpf}, and
the ``Bose Nova'' collapse in Bose gases \cite{donley2002nature}.
Recent experimental developments,
\cite{olsen2004prl_rbk,zaccanti2006pra_rbk,stan2004ofr,ospelkaus:120402,jin2008,ketterle2008}
have enabled the creation of an ultracold mixture of bosons and
fermions, where an interspecies Feshbach resonance may introduce a
rich source of new phenomena. From the theoretical point of view,
studies of Bose-Fermi mixtures to date have been mostly limited to
non resonant physics, focusing mainly on mean field effects in
trapped systems
\cite{roth2002mfi,roth2002sas,modugno2002cdf,hu2003ttb,liu2003fte,
  salasnich2007sbd,pedri2005tdb,
  adhikari:043617,buchler2004psa,kanamoto2006psf,adhikari2005fbs},
phases in optical lattices
\cite{albus2003mba,lewenstein2004abf,roth2004qpa,sanpera2004afb,
  gunter2006bfm,tit2008},or
equilibrium studies of homogeneous
gases, focusing mainly on phonon induced superfluidity
or beyond-mean-field effects
\cite{bijlsma2000ped,heiselberg2000iii,efremov2002pwc,
  viverit2002bis,matera2003fpb,albus2002qft,wang2006sct}.
Pioneering theoretical work on the resonant gas  include Ref.
\cite{powell2005prb}, in which a mean-field equilibrium study of the
gas is supplemented with a beyond-mean-field analysis of the bosonic
depletion; and Ref. \cite{storozhenko2005pra}, where an equilibrium theory
is developed using a separable-potentials model.

The aim of this article is to develop and solve a mean-field theory
describing an ultracold atomic Bose-Fermi mixture in the presence of
an interspecies Feshbach resonance. This goal appears innocuous
enough at first sight, since mean-field theories  for resonant
Bose-Bose \cite{Timmermans1999prl,holland2002prl_ramsey} and
Fermi-Fermi \cite{milstein2002pra,griffin2002prl,stoof2003joptb}
gases exist, and have been studied extensively. In both of these
theories, the mean-field approximation consists of considering the
bosonic Feshbach molecules as being fully condensed, and this
greatly simplifies the treatment, since the Hamiltonian reduces to a
standard Bogoliubov-like integrable form \cite{landau9}.

The fundamental difference between these examples
and the Bose-Fermi mixtures is that in the latter the Feshbach
molecules are fermions, and therefore their center of mass-momentum
must be included explicitly. The most obvious mean-field approach
consists in considering the atomic Bose gas to be fully condensed. However,
 as we will show below, resonant molecules are really composed of two bound atoms,
which spend their time
together vibrating around their center of mass. It follows that
outright omission of the bosonic fluctuations of the atoms, disallows
the bosonic constituents to oscillate (i.e. fluctuate) at all, and
therefore this leads to an improper description of the physics of
atom pairs.

This article is organized as follows. In section \ref{ch3} we
introduce the field theory model used to study Feshbach resonances,
describing briefly the parametrization used, and outlining the exact
solution of this model in the two-body limit. The section ends with
a test of this two-body theory, by comparing the binding and
resonance energies predicted by the model and the virtually-exact
analogues obtained from two-body close coupling calculations.
Section \ref{ch4} introduces the simplest mean-field many-body
theory of the gas, obtained by disregarding all bosonic
fluctuations. The solution of the theory is outlined, and its
limitations highlighted. In spite of these limitations, mean-field
theory provides a useful language for dealing with the problem, the
utility of which will persist even beyond the limits of
applicability of the theory itself.

Finally in section \ref{ch5} we introduce our generalized mean-field
theory, which is, in short, similar to the mean-field theory described
in section \ref{ch4}, but with the notable improvement of using properly
renormalized molecules as building blocks, instead of their bare
counterparts. This approach is not trivially described in the
Hamiltonian formalism, where substituting dressed molecules for free
ones would lead to double counting of diagrams. In this section we
therefore shift to the Green-function/path-integral language, where
this double-counting can be avoided quite easily. Finally we proceed
to the numerical solution of this theory, and note that for narrow
resonances the results are consistent with their mean-field
equivalents. This encourages us to develop a simple theory to study
the molecular formation via magnetic field ramps, and, using an
approach based on the Landau-Zener formalism \cite{lz1,lz2}, we derive analytic
expressions in section \ref{ch6}.

Throughout this article, we work with zero temperature gases, in the free
space thermodynamic limit. These are limitations which render the
results obtained here hard to directly compare with experimental
results. One of the main possible future directions of this work
should include solving the same problem in a trap, and generalization
to higher temperatures.

\section{The Model}
\label{ch3} We are interested primarily in the effects of resonant
behavior  on the otherwise reasonably understood properties of the
system. To this end we use a model which has become standard in the
last few years. This model has been useful in studying the effect of
resonant scattering in Bose
\cite{holland2002prl_ramsey,stoof2004pr,burnett2003pra} and Fermi
\cite{holland2002pra,stoof2004prl,gurarie2004prl,griffin2003pra,sasha2004jpb,sasha2005pra}
gases. In the case of the bose-fermi mixture, this model has been
used in Ref.\cite{powell2005prb},and
\cite{bortolotti2006jpb,bortolotti2006sfm}. We refer to these works
for further details about the origin and justification of the
Hamiltonian we use here, and for details on the solution in the two
body regime.

The Hamiltonian for the system reads:
\begin{equation}
\label{hamiltonian_gen}
H=H_0+H_I,
\end{equation}
where
\begin{eqnarray}
H_0&=&\sum_p \epsilon_p^F \ \hat{a}_p^{\dagger}\hat{a}_p + \sum_p
\epsilon_p^B  \
\hat{b}_p^{\dagger}\hat{b}_p  +  \sum_p \left( \epsilon_p^M + \nu
\right) \
\hat{c}_p^{\dagger}\hat{c}_p \nonumber \\
&+&{\gamma \over 2 V} \sum_{p,p',q}
\hat{b}_{p-q}^{\dagger}\hat{b}_{p'+q}^{\dagger} \hat{b}_{p} \hat{b}_{p'}
\nonumber \\
H_I &=& {V_{bg} \over V} \sum_{p,p',q}
\hat{a}_{p-q}^{\dagger}\hat{b}_{p'+q}^{\dagger} \hat{a}_{p} \hat{b}_{p'}
\nonumber \\
&+&{g \over
  \sqrt{V}}\sum_{q,p}(\hat{c}_q^{\dagger}\hat{a}_{-p+q/2}\hat{b}_{p+q/2} +
h.c.). \nonumber \\
\label{act-bfm}
\end{eqnarray}
Here $\hat{a}_p,\hat{b}_p$, are the
annihilator operators for, respectively, fermions and bosons, $\hat{c}_p$
is the annihilator operator for the molecular field
\cite{holland2002pra,sasha2004jpb,sasha2005pra}; $\gamma = 4\pi a_b /
m_b$ is the interaction term for
bosons, where  $a_b$ is the boson-boson scattering
length; $V_{bg},\nu$, and $g$ are
parameters related to the Bose Fermi interaction, yet to be
determined; the single particle energies are
$\epsilon^{\alpha} = p^2 / 2 m_{\alpha}$, where
  $m_{\alpha}$ indicates the mass of bosons, fermions, or pairs;
and $V$ is the  volume  of a quantization box with periodic boundary
conditions.

The first step is to find the values for $V_{bg},\nu,g$, in terms of
measurable parameters. We will, for this purpose, calculate the 2-body
T-matrix resulting from the Hamiltonian in
Eq. \ref{act-bfm}. Integrating the molecular field out of the real
time path integral,
  leads to the following Bose-Fermi interaction Hamiltonian:
\begin{equation}
H_I^{2body}={1 \over  V} \left( V_{bg} + {g^2 \over E - \nu}\right)
\sum_{p} \hat{a}_{p}^{\dagger}\hat{b}_{-p}^{\dagger} \hat{a}_{p}
\hat{b}_{-p}.
\end{equation}
This expression is represented in center of mass coordinates, and $E$ is the
collision energy of the system. From the above equation we read
trivially the zero energy scattering amplitude in the saddle point
approximation:
\begin{equation}
T = (V_{bg} -{g^2\over\nu}),
\label{t2b}
\end{equation}
which corresponds to the Born approximation (This is akin to
  identifying the scattering amplitude $f=a_{sc}$ in the
  Gross-Pitaevskii equation, where the interaction term would be ${2
  \pi \over m_{bf}} a_{b}$ ). We emphasize that this
approximation is only valid in the zero energy limit, and it does
not, therefore, describe the correct binding energy as a function of
detuning. However, with this approach we obtain an adequate
description of the behavior of scattering length as a function of
detuning, which allows us to relate the parameters of our theory to
experimental observables via the conventional parametrization
\cite{holland2002pra,stoof2004pr}
\begin{equation}
T(B) = {2 \pi \over m_{bf}} a_{bg} \left(1-{\Delta_B
\over (B - B_0)} \right),
\label{e0param}
\end{equation}
where $a_{bg}$ is the value of the scattering length
far from resonance, $\Delta_B$ is the width, in magnetic
field, of the resonance, $m_{bf}$ is the reduced mass, and  $B_0$ is
the field at which the resonance is centered.

The identification of parameters between Eqns., (\ref{t2b}) and
(\ref{e0param}) proceeds as follows: far from resonance, $|B-B_0|
>>\Delta_B$, the interaction is defined by a background scattering
length, via $V_{bg}={2 \pi a_{bg} \over m_{bf}}$. To relate magnetic
field dependent quantity $B-B_0$ to its
energy dependent analog $\nu$, requires defining a parameter
$\delta_B=\partial \nu / \partial B$, which may be thought of as a
kind of magnetic moment for the molecules.
It is worth noting that $\nu$ does not
represent the position of the resonance nor the binding energy of the
molecules, and that, in general $\delta_B$ is a field-dependent
quantity, since the thresholds move
quadratically with field, because of nonlinear corrections to the
Zeeman effect. For current purposes we identify $\delta_B$ by its
behavior far from resonance, where it is approximately constant. Careful
calculations of scattering properties using the model in
Eq. (\ref{act-bfm}), however,
leads to the correct Breit-Wigner behavior of the 2-body T-matrix
\cite{newton}.

Finally we get the following identifications:
\begin{eqnarray}
&V_{bg}={2 \pi a_{bg} \over m_{bf}} \nonumber \\
&g=\sqrt{V_{bg} \delta_B \Delta_B} \nonumber \\
&\nu=\delta_B (B - B_0).\nonumber \\
\end{eqnarray}

\begin{center}
\begin{figure}[ht]
\resizebox{6.3in}{!}{\includegraphics{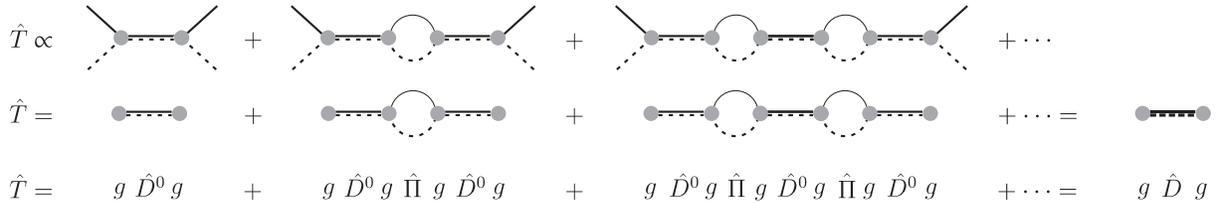}}
\caption{Feynman diagrams representing the resonant collision of a
  fermion and a boson. Solid lines represent fermions, dashed lines
  bosons, and double solid-dashed lines represent the effective
  composite fermions.}\label{2-bdiag1}
\end{figure}
\end{center}

Synthesizing the approach descibed in \cite{bortolotti2006jpb}, and
diagrammatically represented in Fig. \ref{2-bdiag1}, we can
obtain the exact two-body T-matrix of the system by solving the Dyson equation
\begin{eqnarray}
T&=&g D^0 g+g D ^0g\ \Pi \ g D^0 g+g D^0 g\
 \Pi \
g D ^0 g\ \Pi \  g D^0g+\cdots \nonumber \\
&=&g D g
\label{dyson2body}
\end{eqnarray}
where $T$ is the T-matrix for the collision, and which has formal solution
\begin{equation}
T = g D g={g^2\over ( D^0)^{-1} - g^2 \Pi}.
\label{Dyson}
\end{equation}

These quantities take the explicit form
\begin{eqnarray}
& D^0(E) &=\left({V_{bg}\over g^2}+{1 \over E -\nu} \right) \nonumber \\
& \Pi(E)&=-i \int \ {d\omega \over 2 \pi}{d{\bf p} \over (2 \pi)^3}{1
    \over (\hbar \omega - {p^2 \over  2 m_b} + i0^+)}\times \\
  \nonumber
&&{1\over (E- \hbar\omega -
      {p^2 \over
      2 m_f} + i0^+)}   \nonumber \\
& & \approx i{ m_{bf}^{3/2}\over \sqrt{2} \pi} \sqrt{E}+{ m_{bf}
    \Lambda \over \pi^2},
\label{M-E}
\end{eqnarray}
where  $m_{bf}$ is the boson-fermion reduced mass, and $\Lambda$ is
an ultraviolet momentum cutoff needed to hide the unphysical nature
of the contact interactions. Note that $D^0$ represents an effective
molecular field, accounting for the fermion-boson background
interaction, and, as described in detail in
\cite{bortolotti2006jpb}, it is obtained by integrating the original
molecular field, and performing a Hubbard-Stratonovich
transformation \cite{negele} to eliminate the direct boson-fermion
interaction in favour of the effective molecular field $D^0$.

Regularization of the theory
\cite{bortolotti2006jpb,milstein2002pra} is obtained  by the
substitutions
\begin{eqnarray}
&&\bar{V_{bg}}=V_{bg} \left(1 \over 1- {m_{bf} \Lambda V_{bg} \over \pi^2}
\right)\nonumber \\
&&\bar{g}=g \left(1 \over 1- {m_{bf} \Lambda V_{bg} \over \pi^2}
\right)\nonumber \\
&&\bar{\nu}=\nu+\bar{g} g {m_{bf} \Lambda V_{bg} \over \pi^2}.
\end{eqnarray}

Finally the two-body T-matrix takes the form
\begin{equation}
T(E)=\left[ {1 \over V_{bg} +{g^2 \over E - \nu}}+i{ m_{bf}^{3/2}\over
    \sqrt{2} \pi} \sqrt{E} \right]^{-1}.
\label{ch3_tmat}
\end{equation}

\subsection{Poles of the T-Matrix: Testing the Model}
\label{ch3_poles} 
\begin{figure}
\rotatebox{270}{\resizebox{6.in}{6.in}{\centerline{\includegraphics[width=0.75\linewidth,height=0.6\linewidth,
angle=-0]{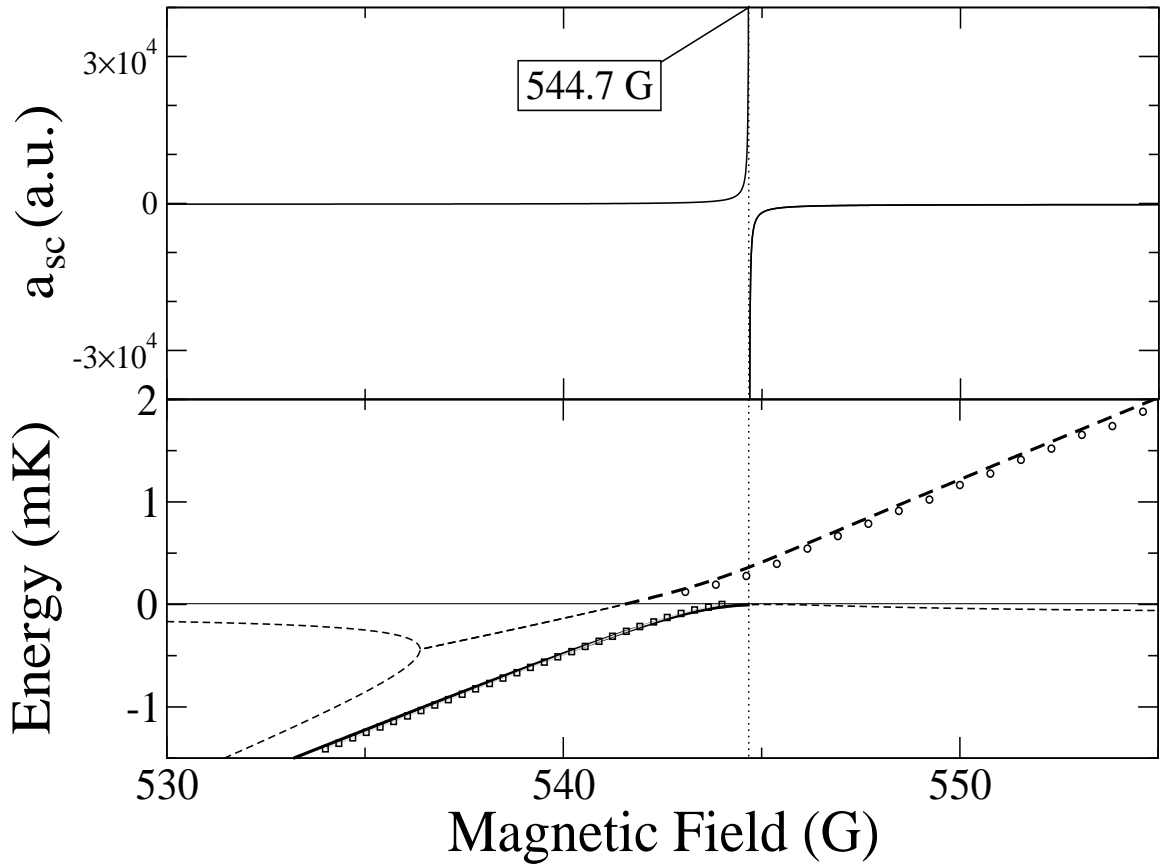}}}} \caption{The top panel shows the scattering
length versus magnetic
  field for the 544.7 G resonance, present in the
  $|9/2,-9/2\rangle|1,1\rangle$
  states of the $ ^{40}$K-$^{87}$Rb collision. The bottom panel shows the
  poles of the model two-body T-matrix
  (Eq. \ref{ch3_tmat}) parameterized for the same resonance,
  as a function of magnetic field.  Thick solid and dashed lines denote the
  real parts of relevant poles, representing bound and resonance
  states respectively. The thin dashed lines are  the real parts of
  unphysical poles. The empty circles and squares, represent the position of
the resonance and the bound state, obtained via a virtually exact
close coupling calculation, and are presented to show the level of
accuracy of the model.}
\label{ch3_twobody_poles}
\end{figure}

Bound states and resonances of the two-body system
are identified in the structure of poles of the T-matrix
(Eq.\ref{ch3_tmat}). This is illustrated in Figure
\ref{ch3_twobody_poles}, where real and imaginary parts of the
poles' energies are plotted as a function of magnetic field. The
resonance portrayed in the figure is the $544.7 G$ resonance present
in the $|9/2,-9/2\rangle|1,1\rangle$  state of  $ ^{40}$K-$^{87}$Rb.
For $B<544.7G$ (corresponding to detunings $\nu<0$) the two-body
system possesses a true bound state, whose binding energy is denoted
by the solid line.  In this case, the pole occurs for real energies.
This bound state vanishes as the detuning goes to zero, where the
resonance occurs.

For positive detunings, $\nu > 0$, on the other hand, the poles are
complex, and
the inverse of the imaginary part is proportional to the lifetime of
the metastable resonant state.
In this regime, there is no longer a true bound
state, but there may be a scattering resonance, indicated in Fig.
\ref{ch3_twobody_poles} by a thick dashed line.  This resonance appears
for magnetic fields $B>544.7$ for this particular resonance, well before
the disappearance of the bound state. This value is highly dependent upon
the value of the background potential. We will see in section
\ref{ch5} that for $V_{bg}=0$, the resonance actually appears at
positive detunings. In the case of $^{40}$K-$^{87}$Rb, $V_{bg}<0$,
implying that there is a weak potential resonance in the open channel
which interferes with the closed-channel resonance, and causes it to
cross the axis at negative detunings. For $V_{BG}>0$
(\cite{unpublished}) the positive
background scattering length is set by a bound state in the open
channel, which does not affect the resonance states, but which
interferes with the bound state at negative detunings.

The thin dashed lines in Fig. \ref{ch3_twobody_poles} are
physically meaningless solutions to the Schr\"{o}dinger equation,
in which the amplitude in the resonant state would grow exponentially
in time, rather than decay.  These poles do not therefore identify
any particular features in the energy-dependent cross section of the
atoms, and will not modify the physics of the system.
Finally Fig.  \ref{ch3_twobody_poles} contains data obtained from
virtually exact close-coupling calculations which show the extent of
validity of the model. For
the purposes at hand this agreement is sufficient.

It should be noted
that the agreement is not as good for positive background scattering
length systems, since the open-channel bound state determining this
scattering length is not adequately described by the model, which
treats the background physics as an essentially zero range
interaction. This implies that the relation between the background
scattering length and open channel bound-state energy is exactly
$E_b=1/2 \mu a_{bg}^2$, while in the physical system this relation
depends on the details of the interaction potential. This problem has
been addressed in the literature \cite{kokkelmans2004pra}, but no
treatable field theory has yet been proposed.

\begin{table}[h!]
\begin{center}
\begin{tabular}{cccc}
\hline
\hline
\\[ 3pt]
$\hspace{.5cm}B_0 (G)\hspace{.5cm}$ & $\hspace{.5cm}\Delta_B (G)
$\hspace{.8cm} &\hspace{.5cm} $\delta_B (K/G) $
\hspace{.5cm}&\hspace{.5cm} $a_{bg} (a.u.)$\hspace{.5cm} \\ [5pt]
\hline
\\[3pt]
$492.49$ & $0.134$ & $3.624\times10^{-5}$ & $-176.5$\\ [5pt]
$544.7$ & $3.13$ & $1.576\times10^{-4}$ & $-176.5$  \\ [5pt]
$659.2$ & $1.0$ & $2.017 \times 10^{-4}$ & $-176.5$\\[8pt]
\hline
\hline
\end{tabular}
\end{center}
\caption{Parametrization of the three main Feshbach resonances used in
  this thesis. All Three resonances are in the $|9/2,-9/2\rangle|1,1\rangle$
  states of the $ ^{40}K-^{87}Rb$ collision}
\label{table_fesh}
\end{table}

\section{Mean-Field Theory}
\label{ch4}
In this section we introduce the many-body physics of the
system, by first analyzing it in a mean-field approach. Because of the
statistical properties of the system, we will see right away that
mean field theory does not recover the correct two body physics in
the low density limits. In spite of this substantial weakness,
however, the approach has several qualitative features which persist
even in the improved theory that we introduce below.
Furthermore, since the model is exactly solvable, it will
allow us to develop a language which will help us to understand the
problem in simpler terms, and to identify some small physical effects,
which, when ignored, can greatly simplify the beyond mean-field
approach presented in the next section.

\subsection{The Formalism}

Starting with the Hamiltonian described by equation
(\ref{hamiltonian_gen}), we obtain the mean-field Hamiltonian by
substituting the boson annihilator $\hat{b}$ by its expectation value
$\phi=\langle \hat{b}\rangle$, a complex number. The number operator
$\hat{b}_p^{\dagger}\hat{b}_p$ therefore becomes $|\phi|^2=N_b$, where
$N_b$ is the number of condensed bosons. The grand canonical
Hamiltonian therefore becomes

\begin{eqnarray}
\label{hamiltonian_mf}
H &=& E_b+\sum_p \left(\epsilon_p^F-\mu_f+V_{bg} n_b\right)
\hat{a}_p^{\dagger}\hat{a}_p \\ \nonumber
 &+&\sum_p \left( \epsilon_p^M + \nu -\mu_m
\right) \
\hat{c}_p^{\dagger}\hat{c}_p \\ \nonumber
&+& g \sqrt{n_b}\sum_{p}\left(\hat{c}_p^{\dagger}\hat{a}_{p} +
h.c.\right),
\end{eqnarray}
where $n_b$ is the density of condensed bosons, $E_b/V = \gamma
n_b^2 -\mu_b n_b$ is the energy per unit volume of the (free)
condensed bosons, a constant contribution to the total energy of the
system, and $\mu_{(b,f,m)}$ are the chemical potentials. These are
Lagrange multipliers
  that serve to keep the densities constant as we minimize the
  energy to find the ground state. In the following we will drop the
  volume term, absorbing it in the definition of the
  creator/annihilator operators, such that the expected value of the
  number operator represents a density, instead of a number.

Before proceeding with the analysis of this Hamiltonian, we
introduce the set of self-consistent equations we wish to solve. To
this end we define the quantities $n_{b(f)}^0$, representing the total
density of bosons (fermions) in the system, at detuning
$\nu \rightarrow \infty$. At finite detunings some of these atoms will combine
into molecules, and the densities will be denoted as $n_{(b,f,m)}$ for
bosons, fermions and molecules respectively.

The system, therefore, is described by six quantities, namely three
densities and three chemical potentials, which require six equations to
determine. These equations, which can be derived by number-conservation constraints
and energy minimization arguments, are:
\newcounter{eqsc}
\stepcounter{equation}
\setcounter{eqsc}{\value{equation}}
\setcounter{equation}{0}
\renewcommand{\theequation}{\arabic{eqsc}.\alph{equation}}
\begin{eqnarray}
&& n_f+n_m-n_f^0=0 \label{eqsc1}\\
&& n_b+n_m-n_b^0=0\label{eqsc2}\\
&& n_f={d \ \Omega \over d\mu_f} \label{eqsc3}\\
&& n_m={d \ \Omega \over d\mu_m} \label{eqsc5}\\
&& {d \Omega \over d \phi}=0 \label{eqsc4}\\
&& \mu_b+\mu_f=\mu_m \label{eqsc6},
\end{eqnarray}
\renewcommand{\theequation}{\arabic{equation}}
\setcounter{equation}{\arabic{eqsc}} where $\Omega=\langle
H\rangle/V$ is the Gibbs free energy. Equations (\ref{eqsc1}) and
(\ref{eqsc2}) follow from the simple counting argument that for
every molecule created, there is one less free boson and one less
free fermion in the gas. Equations \ref{eqsc3}, and \ref{eqsc5} are
simply the Lagrange multiplier constraint equations, equation
(\ref{eqsc4}) follows from the mean-field approximation, whereby the
bosonic field is simply a complex number, and minimization of the
energy can therefore be done directly. Finally equation
(\ref{eqsc6}) is the law of mass action, which follows from the fact
that to make a molecule it takes one free atom of each kind.

The next step is to write down $\Omega$ for the system, by taking the
expectation value of the Hamiltonian in equation( \ref{hamiltonian_mf}),
obtaining
\begin{eqnarray}
\Omega &=& E_b/V+\sum_p \left(\epsilon_p^F-\mu_f+V_{bg} n_b\right)
\eta_f(p)
\\ \nonumber
 &+&  \sum_p \left( \epsilon_p^M + \nu -\mu_m
\right)\eta_m(p) \\ \nonumber
&+& 2 g \sqrt{n_b}\sum_{p}\eta_{mf}(p),
\end{eqnarray}
where $\eta_f(p)=\langle \hat{a}_p^{\dagger}\hat{a}_p\rangle$ and
$\eta_m(p)=\langle \hat{c}_p^{\dagger}\hat{c}_p\rangle$ are the
fermionic and molecular momentum distributions,
$\eta_{mf}(p)=\langle \hat{c}_p^{\dagger}\hat{a}_p\rangle$ is an
off-diagonal correlation term arising from the interactions in the
system, and the densities are given by $n_{f,m,mf}=\int {dp \over 2
  \pi^2} \eta_{f,m,mf}(p)$.
Equations \ref{eqsc1}-\ref{eqsc6} then read
\newcounter{2eqsc}
\stepcounter{equation}
\setcounter{2eqsc}{\value{equation}}
\setcounter{equation}{0}
\renewcommand{\theequation}{\arabic{2eqsc}.\alph{equation}}
\begin{eqnarray}
&& n_f+n_m-n_f^0=0 \label{2eqsc1}\\
&& n_b+n_m-n_b^0=0\label{2eqsc2}\\
&& n_{f,m,mf}=\int {dp \over 2\pi^2} \eta_{f,m,mf}(p) \label{2eqsc3} \\
&& g\ n_{mf}-\mu_b \sqrt{n_b}+\gamma n_b^{3/2}=0 \label{2eqsc4} \\
&& \mu_b+\mu_f=\mu_m \label{2eqsc5}
\end{eqnarray}
\renewcommand{\theequation}{\arabic{equation}}
\setcounter{equation}{\arabic{2eqsc}}
The remaining task is now to find expressions to calculate the expected values
$\eta_{f,m,mf}(p)$. To this end we follow a Bogoliubov-like approach,
similar to that
described in \cite{powell2005prb}. The mean-field Hamiltonian
is bilinear in all creation/annihilation operators, which means that
it can be diagonalized via a change of basis, whereby introducing the
operators
\begin{eqnarray}
\label{change_basis}
\hat{\alpha}_p=A_{\alpha} \hat{a}_p+ C_{\alpha} \hat{c}_p \\ \nonumber
\hat{\beta}_p=A_{\beta} \hat{a}_p+ C_{\beta} \hat{c}_p,
\end{eqnarray}
for some appropriately chosen coefficients $A_{\alpha,\beta}$ and
$B_{\alpha,\beta}$, the Hamiltonian will read
\begin{equation}
H'=E_0+\sum_p \lambda_{\alpha}(p) \hat{\alpha}_p^{\dagger}\hat{\alpha}_p+\sum_p \lambda_{\beta}(p)\hat{\beta}_p^{\dagger}\hat{\beta}_p.
\end{equation}
At this point  we note that the Hamiltonian is just a separable sum of
 free-particle Hamiltonians, where the free particles are fermions,
 with dispersion relations $\lambda_{\alpha,\beta}(p)$. We can readily
 write down the distribution
\begin{equation}
 \eta_{\alpha,\beta}(p)=\Theta(-\lambda_{\beta,\alpha}(p))
\label{etaalphabeta}
\end{equation}
 where $\Theta$ is the step function, and
 calculate the densities $n_{\alpha,\beta}$. The step function could
 be replaced by the free Fermi distribution for non-zero temperatures,
 but the mean-field assumption that all bosons are condensed would no
 longer hold. If these were
 ordinary free fermions with dispersion $p^2/2m-\mu$, equation
 (\ref{etaalphabeta}) would reduce to the standard zero-temperature
 Fermi distribution. We will see below that $\lambda_{\alpha,\beta}(p)$
 are dispersion relations of quasi-particles that are a mixture of
 atoms and molecules.

 Below we show how these ideas, together with equation (\ref{eqsc1}) -
 \ref{eqsc6}, give us the tools we require to
 calculate the observable atomic and
 molecular densities as a function of the chemical potentials.

 To illustrate more explicitly the
diagonalization procedure we define the vectors
\begin{equation}
A=\left(\begin{array}{r} \hat{a}_p \\ \hat{c}_p \end{array}\right)\ \ \
A^{\dagger}=(\ \hat{a}_p^{\dagger} \ \ \hat{c}_p^{\dagger}\ ),
\end{equation}
and
\begin{equation}
B=\left(\begin{array}{r} \hat{\alpha}_p \\ \hat{\beta}_p \end{array}\right)\ \ \
B^{\dagger}=(\ \hat{\alpha}_p^{\dagger} \ \ \hat{\beta}_p^{\dagger}\ ),
\end{equation}
whereby the Hamiltonian can be written as $A^{\dagger}\hat{H} A$, and
$B^{\dagger}\hat{H'} B$,
where
\begin{equation}
\hat{H}=\left( \begin{array}{cc} \left(\epsilon_p^F-\mu_f+V_{bg}
    n_b\right) & g \sqrt{n_b} \\   g \sqrt{n_b} & \left( \epsilon_p^M +
    \nu -\mu_m \right) \end{array} \right),
\label{H_matrix}
\end{equation}
and
\begin{equation}
\hat{H'}=\left( \begin{array}{cc}  \lambda_{\alpha}(p) & 0\\   0 & \lambda_{\beta}(p) \end{array} \right).
\end{equation}
Diagonalizing $\hat{H}$, we get the two eigenvalues
\begin{eqnarray}
\lambda_{\alpha,\beta}(p)&=&{h_f(p)+h_m(p)\over 2} \\ \nonumber
&\pm& {1 \over 2}\sqrt{4 g^2 n_b+(h_m(p)-h_f(p))^2},
\end{eqnarray}
where we have defined $h_f(p)=\left(\epsilon_p^F-\mu_f+V_{bg}n_b\right)$,
and $h_m(p)=\left( \epsilon_p^M +\nu -\mu_m \right)$,
and the unitary eigenvector matrix
\begin{equation}
U=\left( \begin{array}{cc} A_{\alpha} & B_{\alpha} \\ A_{\beta}
    & B_{\beta} \end{array} \right).
\end{equation}
The transformation in eq. \ref{change_basis} can then be written as
$A=U^{\dagger}B$, and its inverse $B=UA$.

Our goal now is to write the densities $\eta_{m,f,mf}(p)$ in terms of the
known densities $\eta_{\alpha,\beta}(p)$. In component notation,
(where $A_i= \hat{a}_p$, etc.), we can write
\begin{equation}
\langle A_i^{\dagger}A_l\rangle=\langle B_j^{\dagger}U_{ji}(U^{\dagger})_{lj}B_k\rangle=U_{lj}U^*_{ij}\langle B_j^{\dagger}B_j\rangle,
\end{equation}
where we have used the fact that since the Hamiltonian is diagonal in
the $B$ basis, then
$\langle B_j^{\dagger}B_k\rangle=\langle B_j^{\dagger}B_j\rangle\delta_{jk}$.

Using this formalism we obtain the relations:
\begin{eqnarray}
\eta_f(p)&=&|A_{\alpha}|^2 \eta_{\alpha}(p)+|B_{\alpha}|^2
\eta_{\beta}(p)\\ \nonumber \eta_f(p)&=&|A_{\beta}|^2
\eta_{\alpha}(p)+|B_{\beta}|^2 \eta_{\beta}(p)\\ \nonumber
\eta_{fm}(p)&=&A_{\alpha}^*A_{\beta}\eta_{\alpha}(p)+B_{\alpha}^*B_{\beta}\eta_{\beta}(p).
\label{self-cons}
\end{eqnarray}
Using these expressions in conjunction with
eqs. \ref{2eqsc1}-\ref{2eqsc5}
will then allow us to compute the equilibrium properties of the system.

\begin{center}
\begin{figure}[h!]
\resizebox{6.5in}{!}{\includegraphics{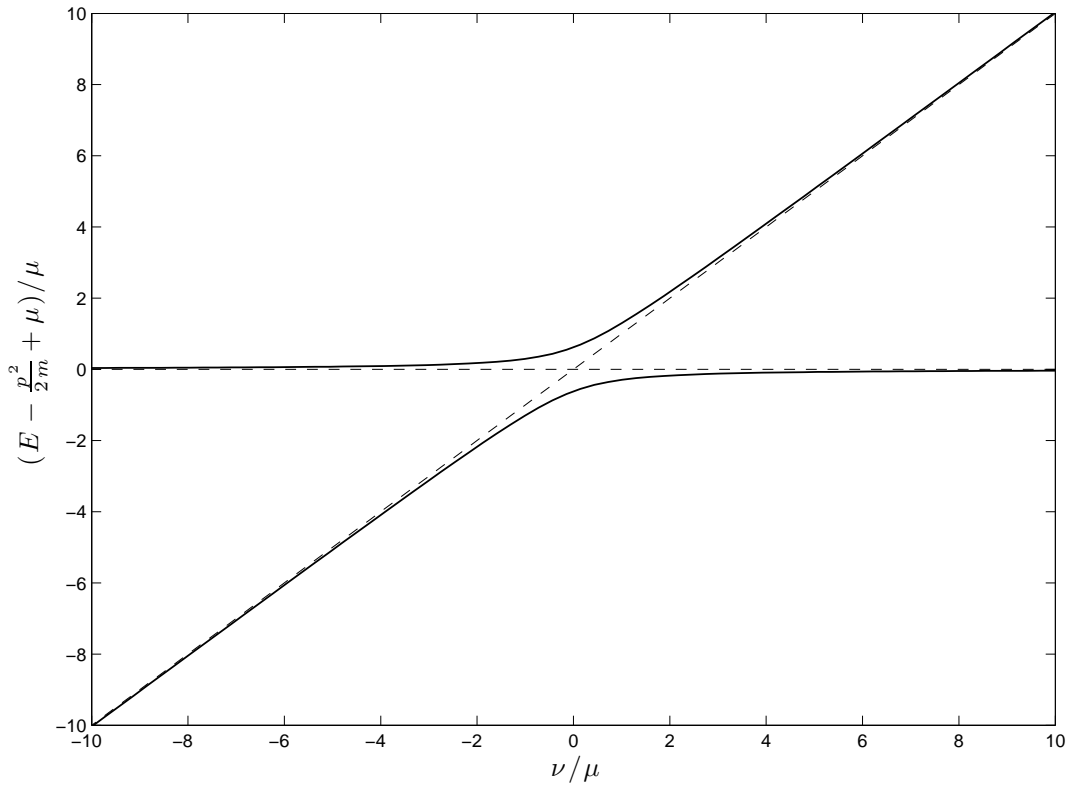}} \caption{ The thick
solid lines represent the ``renormalized'' mean-field energy levels
$\lambda_{\alpha,\beta}(k_f)$ , while the thin dashed  lines
represent their bare counterparts. \label{mf_poles}}
\end{figure}
\end{center}

This simplified, mean-field version of the solution can only approximately
reproduce the energies of atomic and molecular states, as is shown in figure
\ref{mf_poles}.  In this example, we have assumed a uniform mixture of
$^{40}$K and $^{87}$Rb atoms with densities $8.2\times 10^{14}$ cm$^{-3}$
and $4.9 \times 10^{15}$ cm$^{-1}$, respectively.  These densities correspond
to the central density of each species assuming it is confined to a
$100$ Hz spherical trap.  Far from the resonance, these energies asymptote
to zero (representing the atomic state) and to the detuning (representing the
bare molecular state that went into the theory).  Near zero detuning,
these levels cross owing to the coupling term $g \sqrt{n_b}$ in
Eqn.~(\ref{H_matrix}).  The size of this crossing is therefore larger
the larger the bosonic density is. In the crossing region the eigenstates
do not clearly represent
either atoms or molecules, but linear combinations of the two.

\subsection{Mean field: noninteracting case}

\label{emf_analres}

To better understand the structure of the mean-field theory, in this
section we detail its results for a {\it noninteracting} gas, by
setting $g = V_{bg} = \gamma = 0$.  We contrast two different physical
regimes, based on the ratio of bosons to that of fermions,
$r_{bf} \equiv n_b/n_f$.

\begin{figure}[h!]
  \resizebox{6.5in}{6.5in}{\includegraphics{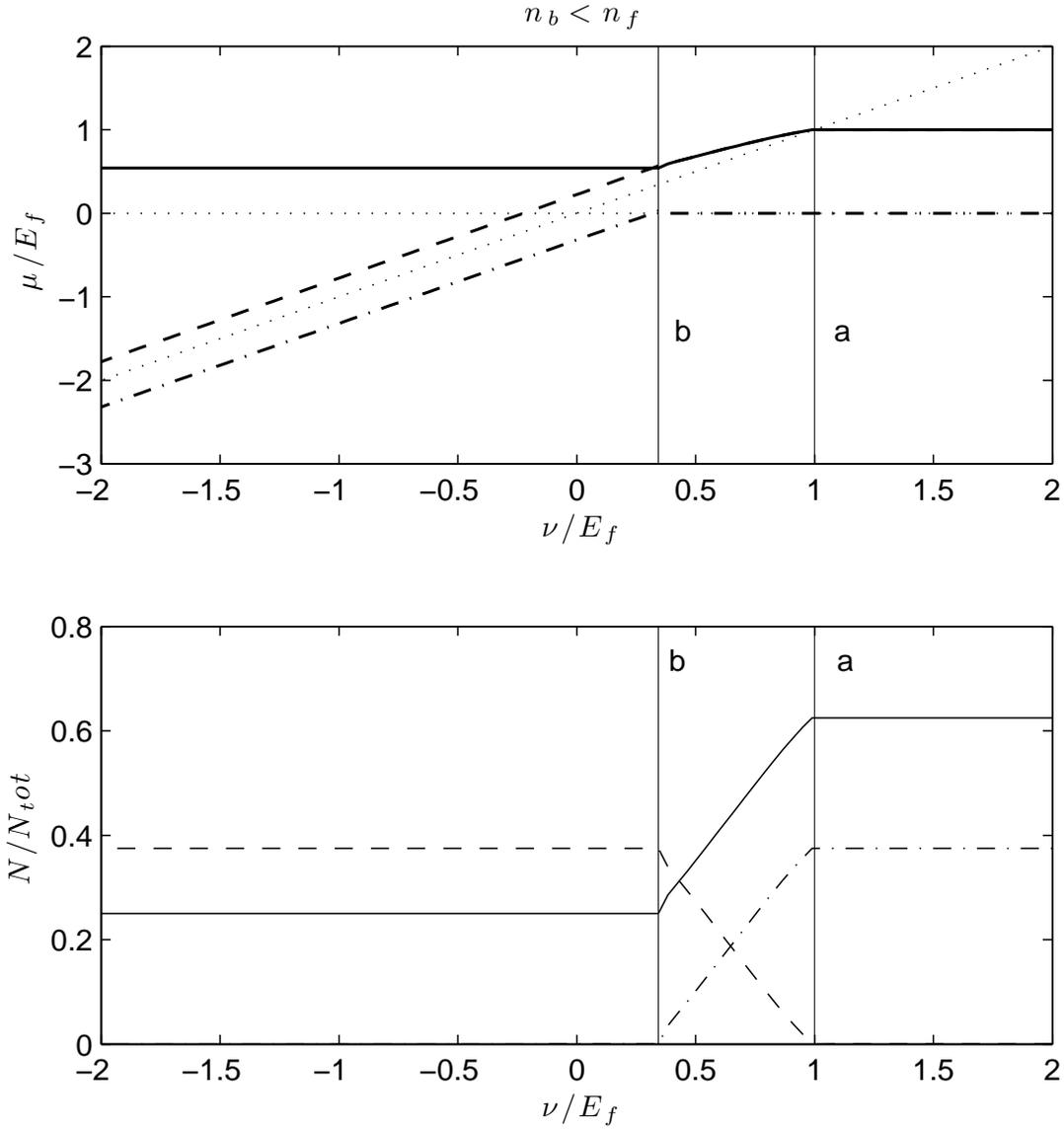}}
\caption{Equilibrium chemical potentials (top panel) and populations (bottom
panel) as a function of detuning for a non interacting
gas with $r_{bf}=.6$. The solid lines represent fermions, dashed lines
molecules, and dashed-dotted lines bosons. The dotted lines in the top
panel represent the bare molecular and fermionic internal energies,
respectively $\nu$ and $0$. The vertical lines labeled a) and b) are
discussed in the text, and represent the detuning at which molecular
formation begins and ends, respectively
\label{mu_free1}}
\end{figure}
\begin{figure}[h!]
\resizebox{6.5in}{6.5in}{\includegraphics{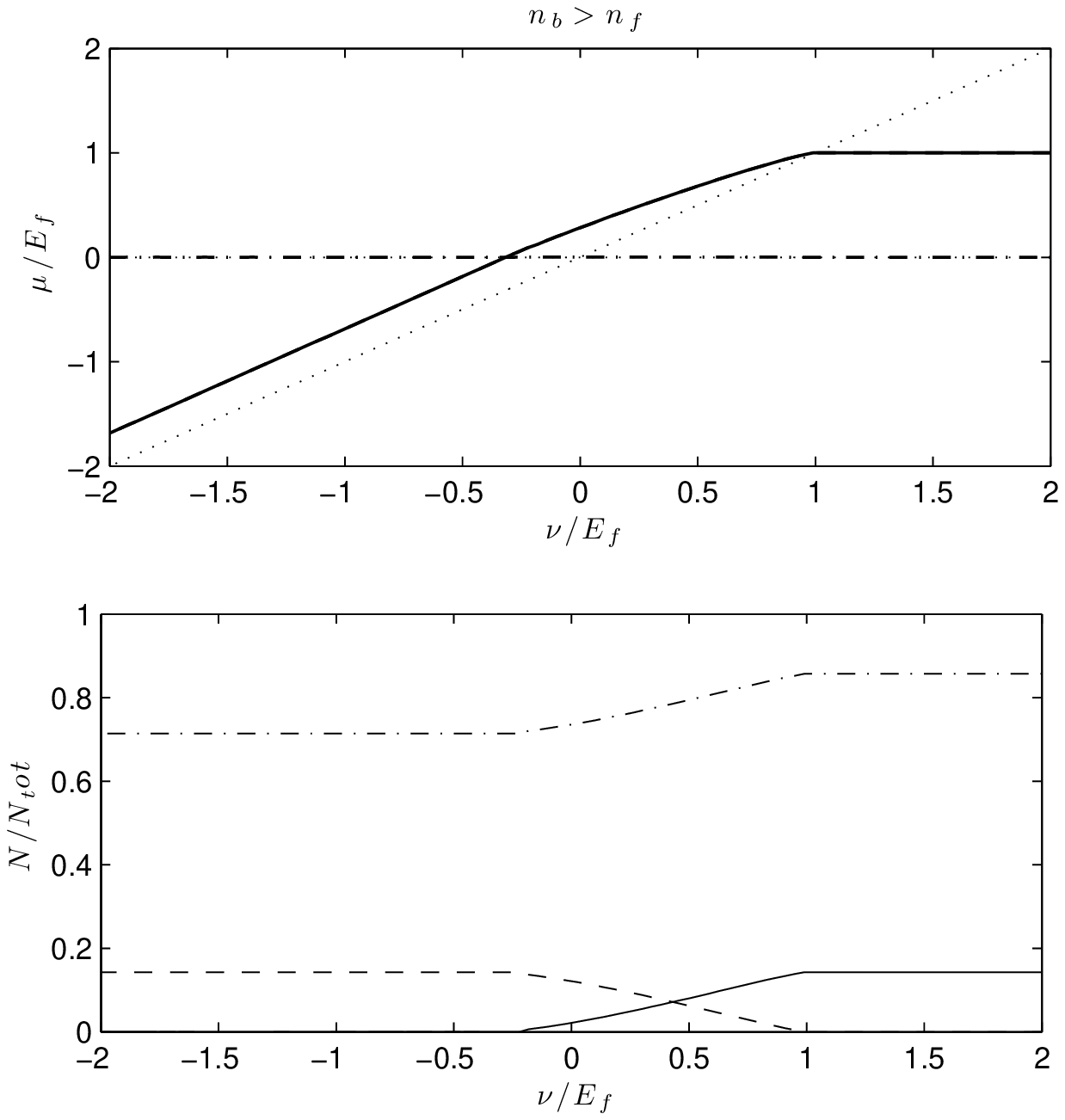}}
\caption{
Equilibrium chemical potentials (top panel) and populations (bottom
panel) as a function of detuning for a non interacting
gas with $r_{bf}=6$. The solid lines represent fermions, dashed lines
molecules, and dashed-dotted lines bosons. The dotted lines in the top
panel represent the bare molecular and fermionic internal energies,
respectively $\nu$ and $0$.
\label{mu_free2}}
\end{figure}

In the case of high fermion density, we
set $r_{bf} = 0.6$, and plot chemical potentials and populations of the various
states in Figure \ref{mu_free1}.  Consider what happens in an infinitely
slow ramp from positive detuning (no molecules) to negative detuning
(introducing bound molecular states).
For large positive detuing, the fermionic and
molecular chemical potentials are the same, since the chemical potential
of the condensed bosons vanishes.  For large enough detuning, the
molecular chemical potential remains below the detuning, so it is energetically
unfavorable to make molecules.  When the detuning dips below the chemical
potential (detuning a in the figure), fermions begin to pair with
bosons to make molecules (see populations in lower panel).  This process
continues until all the bosons are consumed (detuning b), at which point
the populations stabilize.  For detunings less that this, there remain
both fermions and fermionic molecules in the gas, and there are
two Fermi surfaces present.  Because the internal energy of the molecules
continues to diminish at lower detuning, so does the molecular chemical
potential.  The two Fermi surfaces therefore split from one another,
although the relative population of the two fermions is fixed.

By contrast, the case where the bosons outnumber the fermions is
shown in Figure \ref{mu_free2}, where we have set the density ratio
to $r_{bf} = 6$. As in the previous case, no molecules are generated
until the detuning drops lower than the chemical potential of the
atomic Fermi gas. Since there are enough bosons to turn all the
fermions into molecules, there there are no fermionic atoms at
sufficiently negative detuning, and the gas possesses only a single
Fermi surface. The chemical potential of the remaining bosons is
still zero, since these bosons are condensed.  Formally, then, the
chemical potential for atomic fermions is negative, meaning that
their formation at negative detuning is energetically forbidden.

\subsection{Mean field: interacting case}

In most experimental circumstances, the density of bosons is larger than
that of fermions, since condensed bosons cluster to the center of the trap,
whereas fermions are kept away by Pauli blocking.  We therefore
focus on this case hereafter, setting the bose and fermi densities to
$n_b = 4.9 \times 10^{15}$ cm$^{-3}$ and $n_f = 8.2\times 10^{14}$ cm$^{-3}$.
The coupling term $g \sqrt{n_b}$ in equation (\ref{H_matrix}) is the
perturbative expansion parameter for the problem, and since it has
units of energy, it must be
compared with the characteristic non-perturbed energy of the gas,
which in this case is $E_f$. Also since in the perturbative expansions
it always appears squared (see eq. \ref{dyson2body}), we can define the
unitless  small parameter for
the system as $\epsilon_{SM}=g^2 n_b/E_f^2=g^2 n_f/E_f^2 r_{bf}$,
For the $492G$ resonance in
table \ref{table_fesh}, we have $\epsilon_{SM}=6.35\times 10^{-2}
r_{bf}$, and since $r_{bf}=6$, the small parameter is of order
$0.1$, appropriate for perturbative treatment.

\begin{figure}[h!]
\centerline{\includegraphics[width=0.9\linewidth,height=0.55\linewidth,angle=-0]{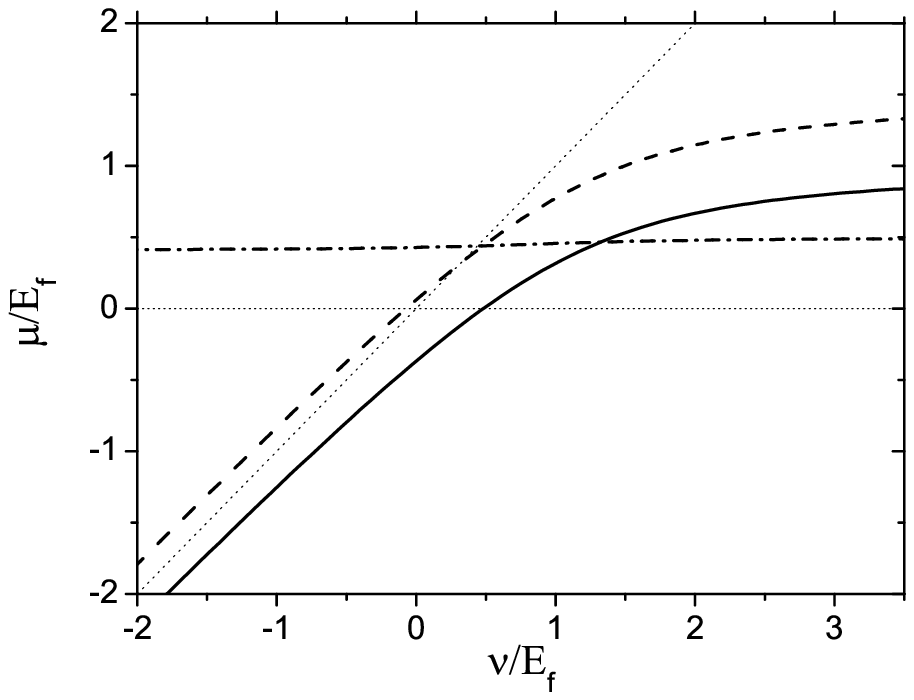}}
\centerline{\includegraphics[width=0.9\linewidth,height=0.55\linewidth,angle=-0]{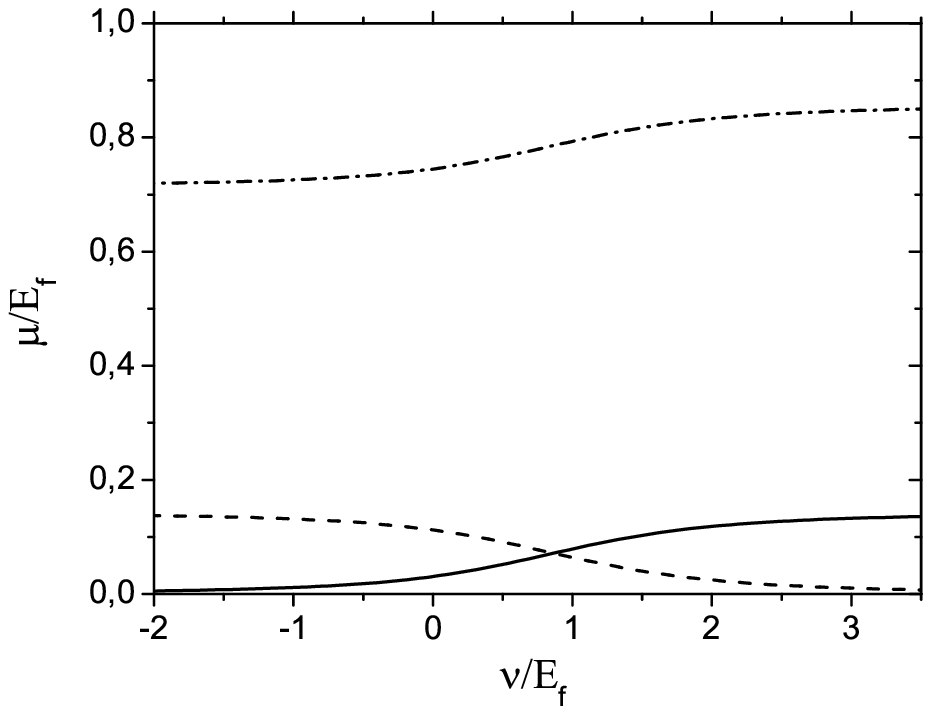}}
 \caption{ Mean-field equilibrium
chemical potentials (top panel) and populations (bottom panel) as a
function of detuning.  The solid lines represent fermions, dashed
lines molecules, and dashed-dotted lines bosons.  \label{mf_mu}}
\end{figure}

Figure \ref{mf_mu}
shows the equilibrium chemical potentials for the system obtained
via a
self-consistent solution of equations (\ref{self-cons}) and
\ref{2eqsc1}-\ref{2eqsc5}.
This figure is qualitatively similar to the corresponding
non-interacting result in Fig.~\ref{mu_free2}, but contains important
differences.  First, the nonzero boson-boson interaction generates
a nonzero bosonic chemical potential $\mu_b$ that
breaks the degeneracy between the molecular and fermionic
chemical potentials.
In physical terms, this means that there is an energy cost in
maintaining bosons unpaired, and therefore we need to take this into
account in the kinematic analysis.  Namely, to make molecules
energetically favorable no longer requires a detuning $\nu$ such that
$\nu = \mu_f$, but now requires $\nu = \mu_f + \mu_b$.  The net result
is to shift the chemical potential up and to the
left by  an amount $\mu_b$, and to shift the molecular population
curve to the left by this amount.

A second difference  is that molecule creation takes place more
gradually as a function of detuning in the interacting case. This is
simply the result of the avoided crossing smearing out the molecular
energy.

Finally, we exploit the simplicity of the mean field approach to
test some approximations that will simplify the beyond-mean-field
approach in the next section.  These approximations have been tested
numerically, and they give corrections of the order of $.1\%$ or
less in calculated molecular populations for all regimes of interest
here.  The approximations are: i) incorporate the boson-boson
interaction $\gamma N_b^3$ by shifting the detuning and chemical
potential as discussed above; ii) disregard the background
scattering between bosons and fermions, i.e., set $V_{bg}=0$, since
this interaction is dominated by its resonance part; and iii)
disregard the correlation function $\langle\eta_bf(p)\rangle$
((analogous to the boson polarization operator in the Green function
formalism), since its contribution to $\mu_b$ is much smaller than
that of $\gamma n_b^3$. It is difficult to directly verify the
validity of these approximations in the beyond-mean-field approach.
Nevertheless, we expect that these approximations remain valid,
since the generalized mean field theory is, after all, a mean-field
theory at heart.

\section{Generalized Mean-Field Theory}
\label{ch5}

In section  \ref{ch4} we reached the conclusion that the mean-field
approach to the resonant Bose-Fermi system does not properly account for the
correct two body physics of the system.
In this section we wish to improve on this, by introducing a
generalization to mean-field theory,  via
an appropriate renormalization of the molecular propagator, which is
able to reproduce the correct two-body physics in the low-density
limit.
To accomplish this, we will have to abandon the Hamiltonian treatment
of the previous section, in favor of a perturbative approach based on
the Green's function formalism, much as was done for two bodies in
\ref{ch3}.   Throughout this section we use the approximations made
above, namely, $\gamma n_b^3 = V_{bg} = \langle \eta_b f(p) \rangle = 0$.

\begin{figure}[h!]
\resizebox{6.3in}{!}{\includegraphics{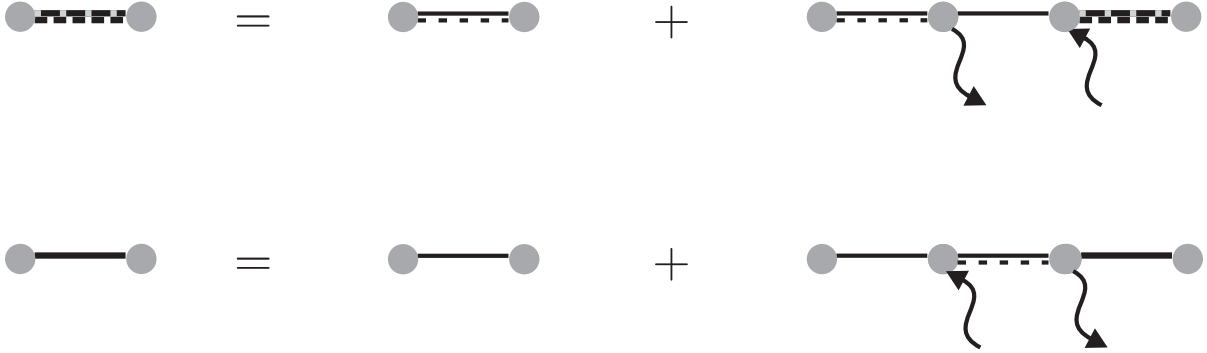}} \caption{
Feynman diagrams included in the mean-field theory. Thin (thick)
solid lines represent free (renormalized) fermions, thin double
dashed-solid lines represent free molecules, and thick double dashed
lines represent renormalized molecules. The little lightning bolts
represent condensed bosons, whereby the arrow indicates whether they
are taken from or released into the condensate. \label{diags_mf}}
\end{figure}

We begin by recasting the rn field result from Sec. \ref{ch4}
in the language of Green functions.
The self-consistent Dyson equations that describe this system are
\begin{eqnarray}
G^{MF}_F(E,P)&=&G^0_F(E,P)+g^2 n_b \ D^0(E,P)\ G^{MF}_F(E,P)\nonumber \\
D^{MF}(E,P)&=&D^0(E,P)+g^2 n_b \ G^0_F(E,P)\ D^{MF}(E,P),
\end{eqnarray}
where the free propagators are simply
\begin{eqnarray}
D^0={1\over \w -\xi^M(p)+i \eta \  {\rm sign}(\xi(p))} \nonumber \\
G^0_{F}={1\over \w -\xi^F(p)+i \eta \  {\rm sign}(\xi(p))},
\label{free_many_body}
\end{eqnarray}
and $\xi^{M,F}(p)=(\epsilon_p^{M,F}-\mu_{m,f})$
These propagators are described diagrammatically in
Fig. \ref{diags_mf}.  They represent the fact that a
free fermion may encounter a {\it condensed} boson and associate
with it, temporarily creating a molecule; or that a free molecule
may temporarily split into a fermion and a condensed boson.
Self-consistency ensures that these
processes may be repeated coherently an infinite number of times.
We neglect the bosonic renormalization equation $\phi^{MF}=n_b^0+g^2 n_b\
G^0_F(E,P)\ D^0(E,P)$, whereby a condensed boson  may
pick-up a fermion to create a molecule; this is equivalent to the
condition $\langle \eta_b f(p) \rangle= 0$.

Solutions to these equations take the form
\begin{eqnarray}
\label{mean_field_green}
G^{MF}_F(E,P)&=&{1\over G^0_F(E,P)^{-1}-g^2 n_b \ D^0(E,P)}\nonumber \\
D^{MF}(E,P)&=&{1\over D^0(E,P)^{-1}-g^2 n_b \ G^0(E,P)}.
\end{eqnarray}
Using the definitions of $G^0_F(E,P)$ and $D^0(E,P)$ from Eq.
\ref{free_many_body}, we can find the poles corresponding to
many-body bound states.  In this case, it can be shown that the
poles are exactly the mean-field eigenvalues
$\lambda_{\alpha,\beta}(p)$ from section \ref{ch3}.  Moreover, the
equations (\ref{mean_field_green}) are symmetric with respect to
interchange of $G_F$ and $D$, which implies that both renormalized
green functions have the same poles, and the same residues. We can
therefore  study the properties of the fermions by only looking at
the molecules. This is not completely surprising, since, given that
the condensed bosons are relatively inert, every molecule
corresponds exactly to a missing fermion, and vice-versa.

The most important deficiency of this mean-field approach is that it
only allows molecules to decay into a free fermion and a condensed
boson, disregarding the possibility that the bosonic byproduct may
be noncondensed. We must allow noncondensed bosons somehow, and yet
these bosons make a perturbation to the result, as seen by the
following argument. The fundamental mean field assumption is that
the gas is at zero temperature, and therefore the noncondensed
population should be negligible at equilibrium. Furthermore, if a
molecule is composed of a zero-momentum boson and a fermion from the
Fermi sea, dissociating into a noncondensed boson implies that the
outgoing fermion would have momentum lower than the Fermi momentum,
an event which Pauli blocking makes quite unlikely. Therefore,  if a
molecule does indeed decay yielding a non-condensed boson, it should
immediately recapture the boson  in a virtual process such as that
described in Fig. \ref{2-bdiag1}. It is only convenient that these
events are exactly the kind of events which will correctly
renormalize the binding energy of the molecules, leading to a theory
which will reproduce the exact two-body resonant physics.

\begin{figure}[h!]
\rotatebox{0}{\resizebox{6.3in}{!}{\includegraphics{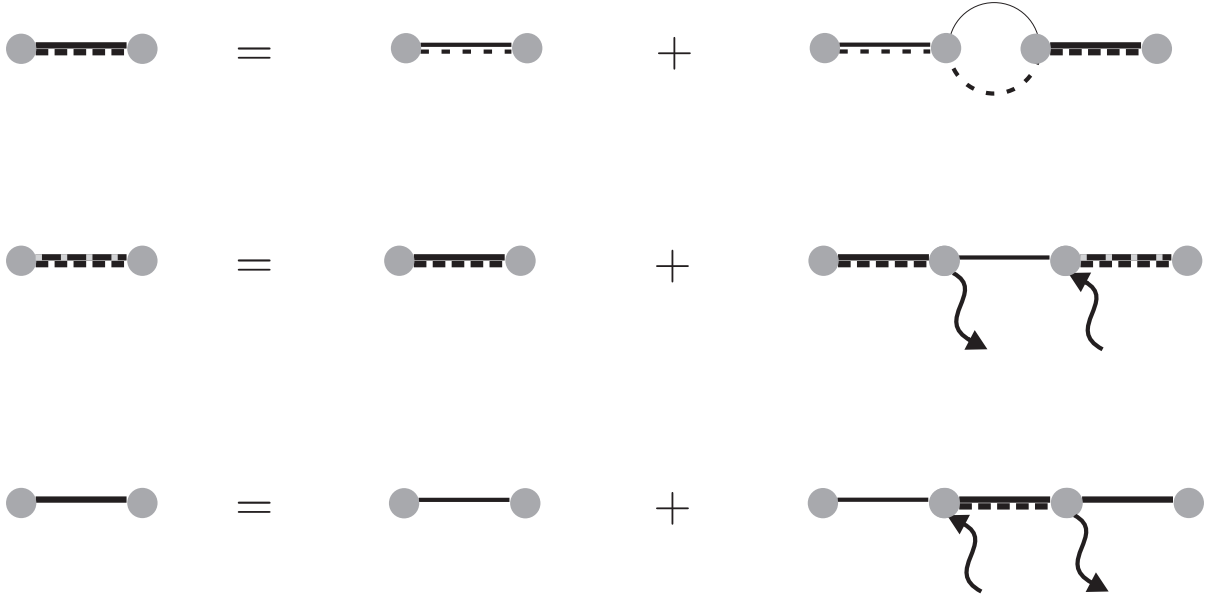}}}
\caption{ Feynman diagrams included in the generalized mean-field
theory. Like in the mean field case (Fig. \ref{diags_mf}), thin
(thick) solid lines represent free (renormalized) fermions, thin
double dashed-solid lines represent free molecules, and thick double
dashed lines represent renormalized molecules. The little lightning
bolts represent condensed bosons, whereby the arrow indicates
whether they are taken from or released into the condensate. The
novelty here in the inclusion of the 2-body dressed  molecules from
Fig. \ref{2-bdiag1}. \label{diags_gmf}}
\end{figure}

\begin{figure}[h!]
\resizebox{6.5in}{6in}{\includegraphics{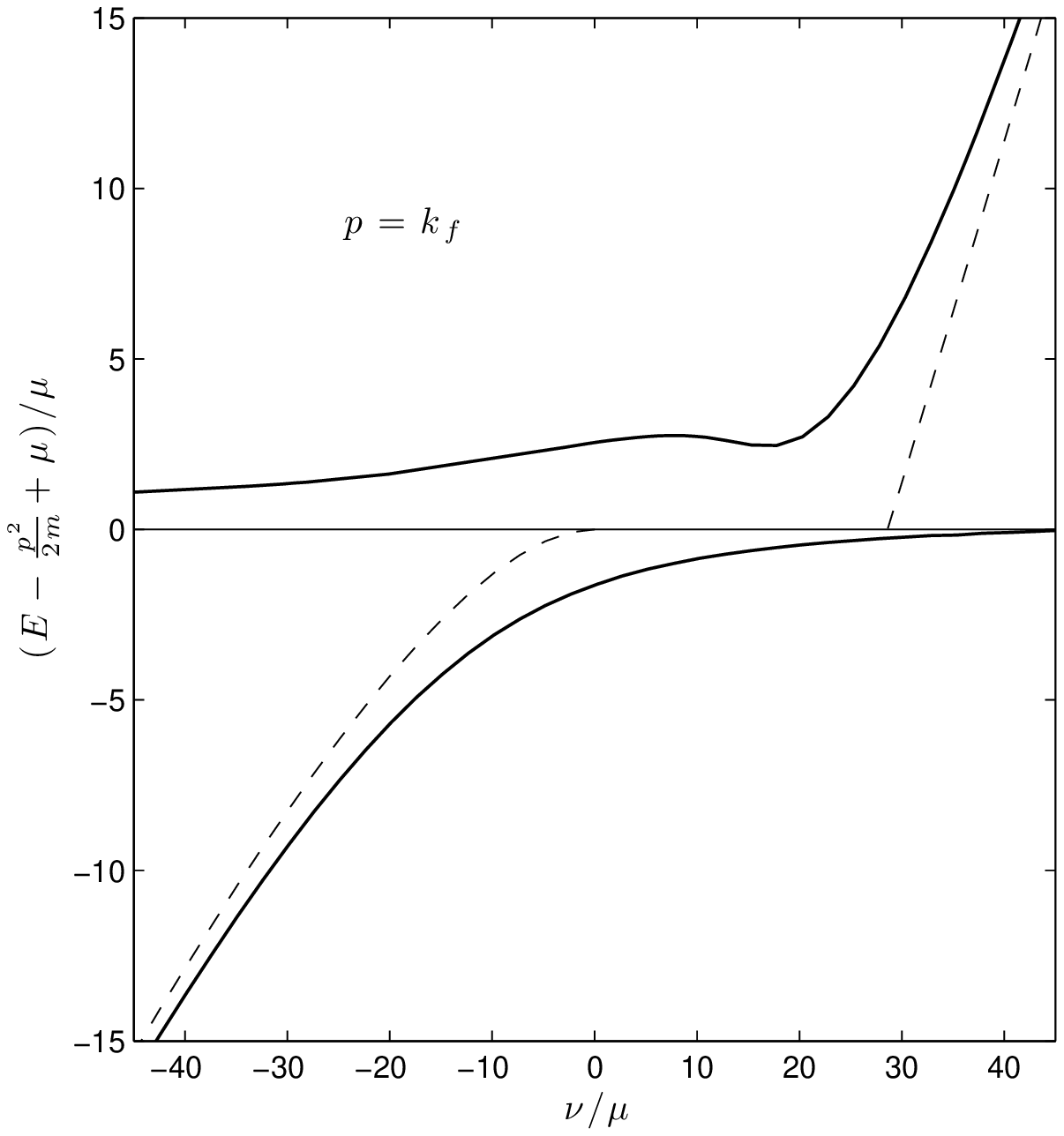}} \caption{The
thick lines represent the poles of $D^{GMF}(k_f)$, and correspond to
molecular and fermionic energies. The dashed lines represent the
molecular two body poles, corresponding to the molecular binding and
resonant energies obtained disregarding the background interaction
\cite{bortolotti2006sfm}. As in the mean-field theory case, the
effect of the interaction with the condensate is to create an
avoided crossing between the atomic and molecular states. The
``bulge'' in the upper solid curve is a consequence of Pauli
blocking which has the consequence of favouring molecular stability.
See \cite{bortolotti2006sfm} for more details.} \label{gmf_poles}
\end{figure}

The Dyson equation describing this generalized mean-field theory are:
\begin{eqnarray}
G^{GMF}_F(E,P)&=&G^0_F(E,P)+g^2 n_b \ D(E,P)\ G^{GMF}_F(E,P)\nonumber \\
D^{GMF}(E,P)&=&D(E,P)+g^2 n_b \ G^0_F(E,P)\ D^{GMF}(E,P).\nonumber \\
\end{eqnarray}
Here we have replaced the free propagator $D^0$ by the renormalized
molecular propagator $D$ from equation
(\ref{Dyson}). A diagrammatic representation of this theory appears
in Fig. \ref{diags_gmf}. By analogy with the mean field version,
 the solution to these equations are:
\begin{eqnarray}
G^{GMF}_F(E,P)&=&{1\over G^0_F(E,P)^{-1}-g^2 n_b \ D(E,P)}\nonumber \\
D^{GMF}(E,P)&=&{1\over D(E,P)^{-1}-g^2 n_b \ G^0(E,P)}.
\end{eqnarray}
These equations preserve the symmetrical nature of the mean-field
theory described above, and also the avoided crossing of atomic and
molecular levels.  This is demonstrated in Figure \ref{gmf_poles},
where we reproduce the $P=k_f$ pole of $D$ from Fig.
\ref{ch3_twobody_poles} as dashed lines.  We also present in this
figure the corresponding poles for the generalized mean-field
theory, as solid lines. As in figure \ref{mf_poles} we note the
splitting in two energy levels, avoiding each-other around $\nu=0$.
The fundamental difference in this case is that the molecular curve
does not asymptote to the bare detuning, but rather to the correct
molecular binding and resonance energies.

Studying the equilibrium properties of the system is now a matter
solving the self-consistent set of equations
(\ref{2eqsc1}-\ref{2eqsc5}), while setting $\eta_{mf}$ and $\lambda$
equal to zero. To do this we first need to extract the distributions
$\eta_{f,m}$ from the Green functions $D^{GMF}$ and $G^{GMF}_F$. To
avoid taking a distracting detour here, we refer the reader to
appendix \ref{appGF} for details. As in the previous section, we
will consider a mixture composed of a free gas of fermionic $^{40}K$
atoms, with a density of $8.2\times 10^{14}$ cm$^{-3}$, and a gas of
condensed $^{87}Rb$ bosons with density $4.9 \times 10^{15}$
cm$^{-3}$ (corresponding to the respective Thomas-Fermi densities of
$10^6$ atoms of either species in the center of a $100Hz$ spherical
trap).

\begin{figure}[h!]
\resizebox{5.5in}{!}{\includegraphics{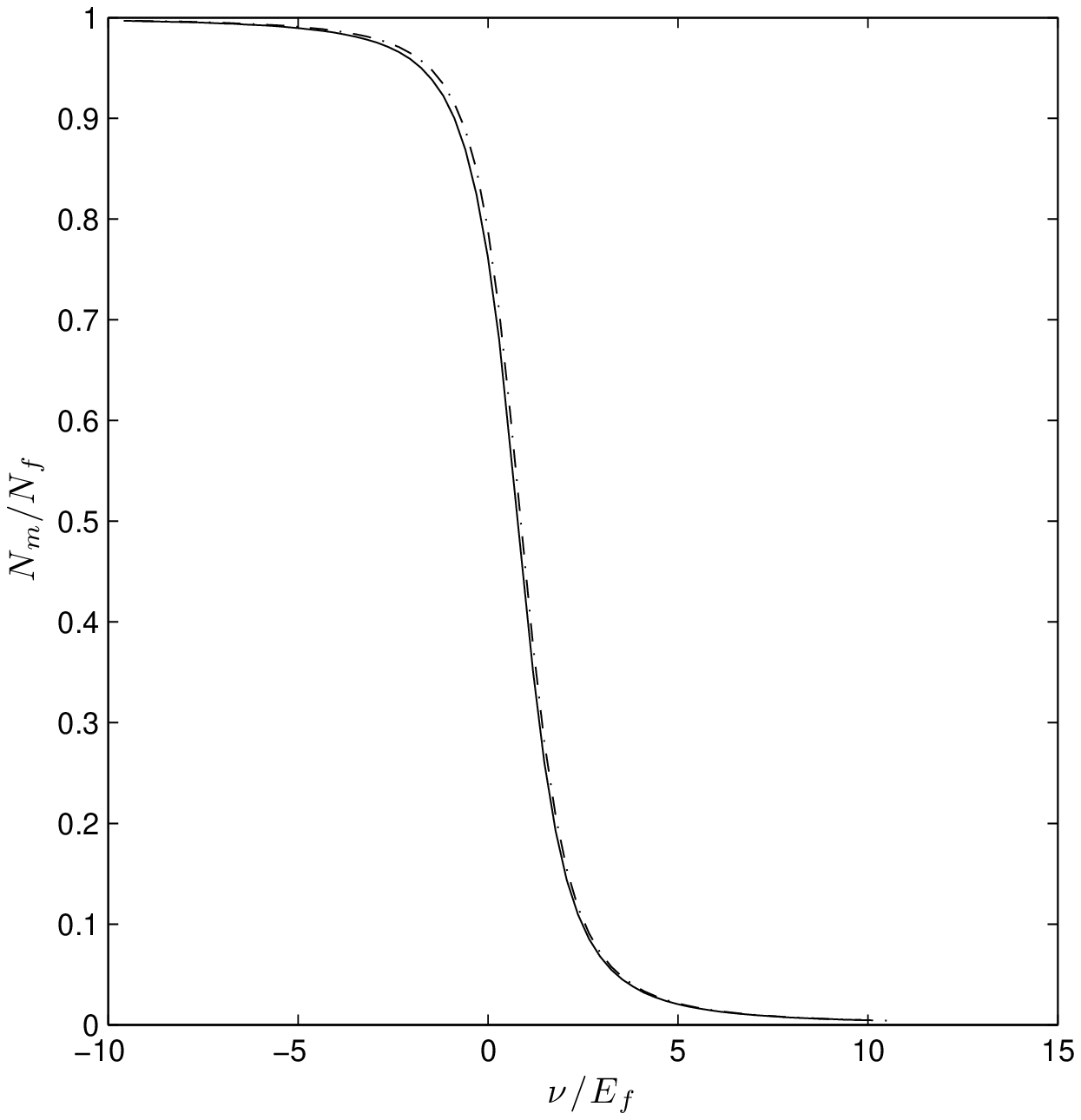}} \caption{
Equilibrium molecular population as a function of  detuning for the
narrow $492.49G$ resonance. The solid line represents results
obtained via the generalized mean field theory presented in the
text, while the dashed-dotted line represents the mean field
results. \label{pop_narrow}}
\end{figure}

\begin{figure}[h!]
\resizebox{5.5in}{!}{\includegraphics{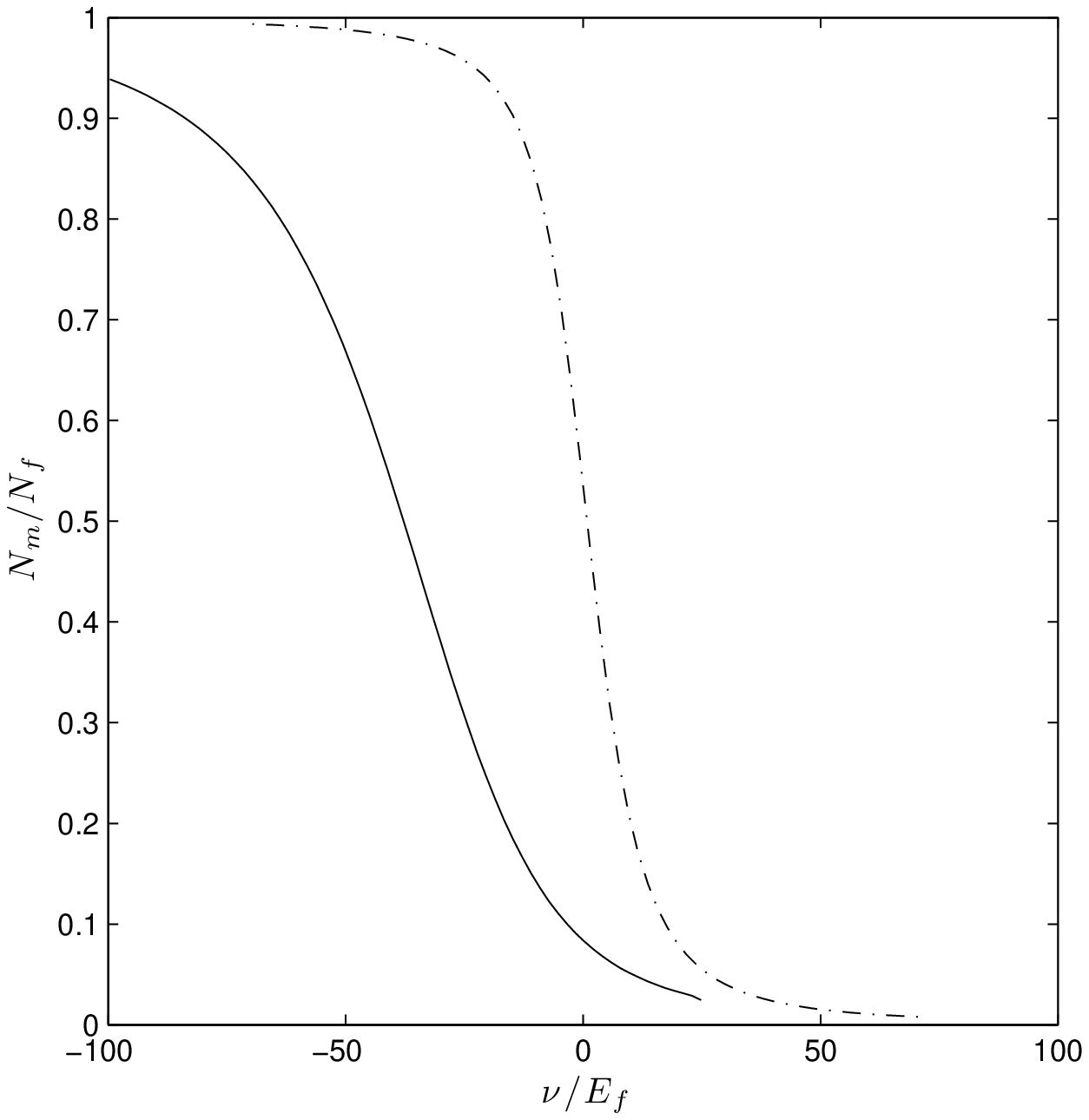}} \caption{
Equilibrium molecular population as a function of  detuning for the
wider $544.7G$ resonance. The solid line represents results obtained
via the generalized mean field theory presented in the text, while
the dashed-dotted line represents the mean field results.
\label{pop_wide}}
\end{figure}

Figure \ref{pop_narrow} shows the equilibrium molecular population
as a function of detuning, for the $492.5G$ resonance. For the
densities assumed,  the mean-field parameter  $\epsilon_{SM}=g^2
n_b/E_f^2\approx 0.4$ is indeed perturbative. For this narrow
resonance, the agreement between mean-field and generalized
mean-field is quite good, and we could as easily have used the bare
molecular positions to calculate this quantity. However, the
situation is completely different for the wide resonance at 544.7G,
for which the equilibrium molecular populations are shown in Fig.
\ref{pop_wide}.  Here the ``perturbative'' parameter has the value
$\epsilon_{SM}=38.7$, and is not perturbative at all. For a given
small detuning, the simple mean-field approximation would greatly
overestimate the number of molecules in the gas at equilibrium.  The
more realistic generalized mean field theory accounts for the fact
that the actual molecular bound-state energy is higher than the bare
detuning.  This fact in turn hinders molecular formation, according
to chemical potential arguments analogous to those in Sec.
\ref{emf_analres}.

\section{Molecule Formation}
\label{ch6} Moving beyond equilibrium properties, we are also
interested in the prospects for molecule creation upon ramping a
magnetic field across the resonance. In reference
\cite{bortolotti2006jpb}, the mean-field equations of motion for the
system at hand were derived, as follows:
\newcounter{eqp6}
\stepcounter{equation}
\setcounter{eqp6}{\value{equation}}
\setcounter{equation}{0}
\renewcommand{\theequation}{\arabic{eqp6}.\alph{equation}}
\begin{eqnarray}
i\hbar {\partial \over \partial t} \phi &=& (V_{bg} \rho_F+\gamma
  |\phi|^2) \phi + g \rho_{MF}^* \label{eqr1} \\
\hbar {\partial \over \partial t} \eta_F(p)&=&-2 g \ \Im m  (\phi
  \eta_{MF}(p))\label{eqr2}\\
\hbar {\partial \over \partial t} \eta_M(p)&=& 2 g \ \Im m  (\phi
  \eta_{MF}(p))\label{eqr3}\\
i \hbar {\partial \over \partial t} \eta_{MF}(p)&=&\left[\epsilon_p^F-
  \epsilon_p^M -\nu+ V_{bg} |\phi|^2\right]
  \eta_{MF}(p) -  \nonumber \\ && \ \ \ \ g
  \phi^*\left(\eta_F(p)-\eta_M(p)\right) \label{eqr4},
\end{eqnarray}
\renewcommand{\theequation}{\arabic{equation}}
\setcounter{equation}{\arabic{eqp6}}
where
$\eta_{F}(p)=\langle \hat{a}_{p}^{\dagger} \hat{a}_{p}\rangle$ is the
  the fermionic distribution, $\eta_{M}(p)$ its molecular counterpart,
  and $\rho_{M,F}=\int {dp \over 2 \pi^2} p^2  \eta_{M,F}(p) $ the fermionic and molecular
  densities. Similarly
 $ \eta_{MF}(p)=\langle \hat{c}_{p}^{\dagger} \hat{a}_{p}\rangle$ and
    is the
   distribution for molecule-fermion correlation, with the
  associated density $\rho_{MF}$.

Reference \cite{bortolotti2006jpb} outlined the limitations of the
non-equilibrium theory, by claiming that to obtain the correct
two-body physics in the low-density limit it would be necessary to
include three point and possibly higher correlations. While this
fact is indeed true, we have amended it in the previous section. For
experimentally reasonable parameters, the mean-field theory can be
complemented by the correct renormalized propagator to accurately
describe the equilibrium properties of the gas. Encouraged by this
argument, we now apply it to the problem of a field ramp as well.

In the following we wish to study molecular formation via a time
dependent ramp of the
magnetic field across the resonance. To this end we use two approaches:
the first consists of propagating equations (\ref{eqr1}-\ref{eqr4}),
ramping the detuning linearly in time from a large positive value to a
large negative one, and plotting the final molecular population as a
function of detuning ramping rate $R$. The second approach consists in
noticing that if $\nu(t)$ is a linear function of time, then
the mean-field Hamiltonian (eq. \ref{H_matrix}) is ideally suited to
a Landau-Zener treatment, whereby the final molecular population as a
function of detuning can be readily written as
\begin{equation}
n_m/min(n_b,n_f)=1-e^{-R \over \tau}
\end{equation}
Here $n_m/max(n_b,n_f)$ is the fraction of possible molecules formed,
and $R=1 / {\partial B \over \partial t}$ is the inverse ramp rate,
and the exponential time constant is given by $\tau={\hbar \delta_B
    \over g^2 n_b}={m_{bf} \over h a_{bg} n_b \Delta_B}$, where
  $a_{bg}$ is the background scattering length, $h$ is Plank's
  constant, and $\Delta_B$ is the magnetic field width of the
  resonance.

Remarkably, the characteristic sweep rate $\tau$ does not
depend on the fermionic
density. This arises from the fact that in mean-field theory
the momentum-states of the fermionic gas are uncoupled, except via
the depletion of the condensate.  Since in
the Landau-Zener approach the depletion is assumed small, it follows
that the various fermionic momentum states are considered
independently, and thus the probability of transition of the gas is
equal to the probability of transition of each individual momentum
state. This approximation is only valid for narrow
resonances, such as the $492.5G$ resonance in table
\ref{table_fesh}.

\begin{figure}[h!]
\resizebox{6.5in}{!}{\includegraphics{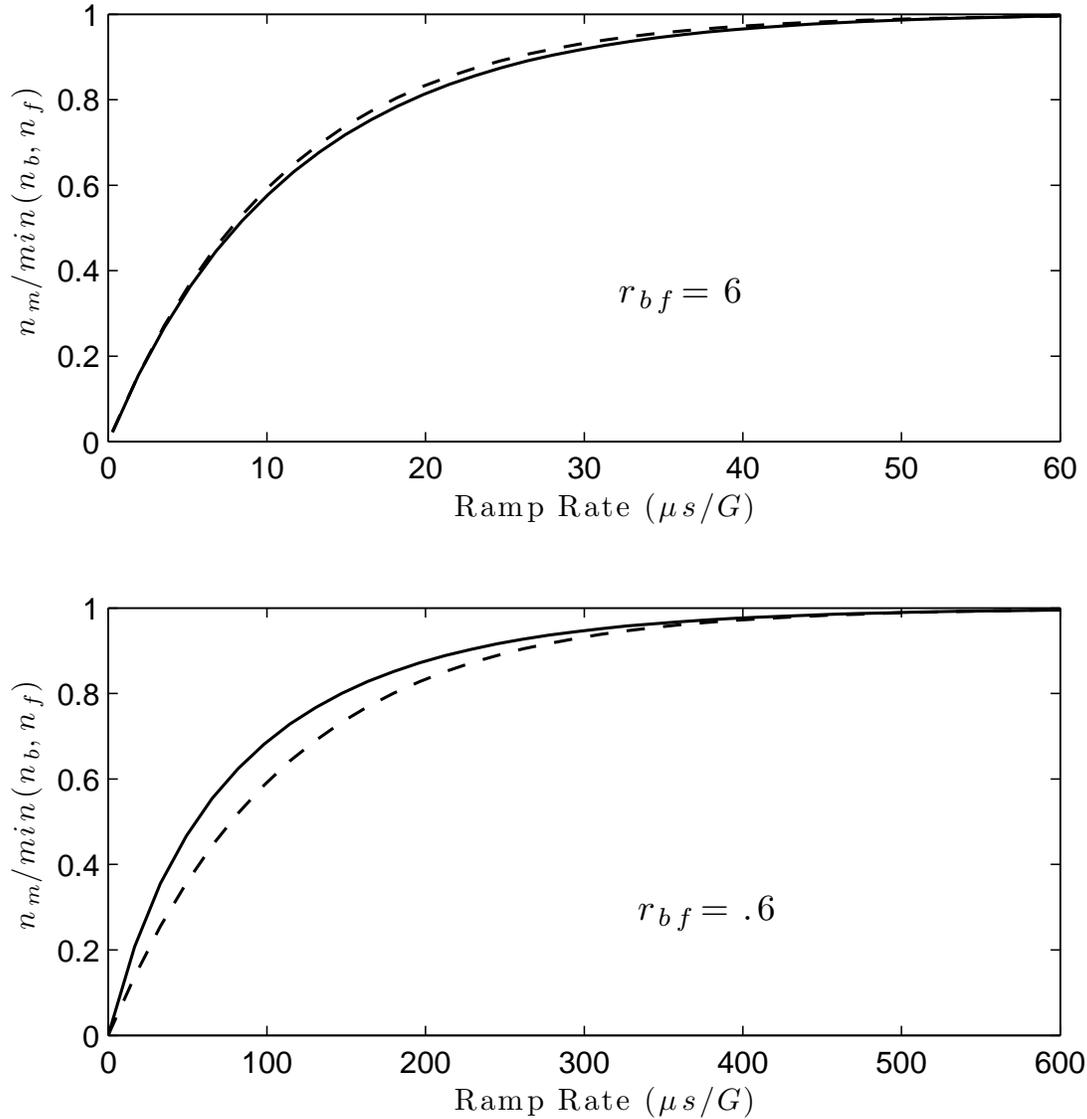}} \caption{ Transition
probability into molecular state via a magnetic field ramp across
the  $492.5G$ Feshbach resonance. The solid lines are obtained via
numerical solutions of equations (\ref{eqr1}-\ref{eqr4}), while the
dashed lines represent the Landau-Zener equivalent. In the top panel
the gas is composed of more bosons than fermions $r_{bf}=6$, while
in the bottom panel the opposite is true $r_{bf}=.6$ \label{LZ}}
\end{figure}

Molecular formation rate versus ramp rate is shown in Figure \ref{LZ},
for the cases $r_{bf}=6$ (more fermions than bosons) and
$r_{bf}=0.6$ (more fermions than bosons.  In both cases
the Landau-Zener result agrees nearly perfectly with direct
numerical integration.  This agreement is surprising in the
case of more fermions, since
the width of the crossing is proportional to
density of leftover bosons, and we expect that this number will change
substantially as the bosonic population is depleted via the formation
of molecules. This type of time dependent crossing should not be
properly described by the Landau-Zener formula. However, monitoring
the time evolution of the molecular population as a function of time
shows that the majority of the transfer takes place quite abruptly
somewhat after crossing the zero-detuning region, whereby the change
in bosonic density does not modify the energy levels
substantially.

\section{Conclusion}
In this article, we developed and solved a generalized mean-field theory
describing an ultracold atomic Bose-Fermi mixture in the presence of
an interspecies Feshbach resonance. The theory is ``generalized''
in the sense that it correctly incorporates bosonic fluctuations,
at least to the level that it reproduces the correct two-body
physics in the extreme dilute limit.
This theory, like any mean-field theory, presents undeniable
limitations. Nevertheless, any useful
many-body treatment must start from a well conceived mean-field
theory.
Future directions of this work
should include the generalization to finite temperature, and the
inclusion of a trap, initially in a local-density approximation. These
advances would be essential to check for empirical confirmation
of the theory.

\acknowledgements

DCEB and JLB acknowledge support from the DOE and the Keck Foundation.

\appendix
\renewcommand{\theequation}{A-\arabic{equation}}
\section{Green Function Methods for Fermions}
\label{appGF}
In this Appendix we briefly introduce some of the Green function techniques
that we found useful in our calculations.

\subsection{Free Green functions}
We start from the Green function for a gas of free fermions, which is
given, in the frequency-momentum representation by
\begin{equation}
G^0(w,\bq)={1\over \w -\xi(\bp)+i \eta \  {\rm sign}(\xi(\bp))},
\end{equation}
where $\xi(\bp)=\bp^2/2 m -\mu$.
The momentum distribution, at equilibrium, is given by
\begin{equation}
n(\bp)=-  i \lim_{\eta \to 0^+} \int{{d\w \over 2 \pi} e^{i \w
    \eta}G^0(w,\bq)},
\label{free_num_int}
\end{equation}
Here the limit comes from the equilibrium condition; the frequency,
in the Green function definition, is the fourier space equivalent of
time, whereby the real time green function represents the evolution
of the system from  from time $t$ to $t'$, and the observables
obtained this way, represent expected values of the kind $ \langle
\psi(t)|\mathcal{O}|\psi(t')\rangle$. However, since we want
equilibrium conditions, we need to take the limit $t\to t'$, which
is non trivial, since $G^0$ is defined by a green function equation
of the form $\mathcal{L}\ G^0(t-t')\propto \delta(t-t')$, where
$\mathcal{L}$ is  some linear operator, and which highlights a
peculiar behavior in the limit we desire. However, since we know on
physical grounds that observables, such as the momentum
distribution, must be defined and well behaved at equilibrium, then
by first taking the expectation value integral, and then the limit,
we can circumvent the problem. In eq. \ref{free_num_int} this
implies we cannot quite get rid of  the fourier transform exponent
$e^{i \w (t-t')}$ until after the $\w$ integral.

To perform the integral in \ref{free_num_int}, we exploit the fourier
exponent, by noting that since $\eta$ is positive, $e^{i \w \eta}
\to 0$ as $\w \to +i \infty$, so that the integral is identical to a
contour integral over the path defined by the real $\omega$ axis,
closed in the upper complex $\omega$ plain by an infinite radius
semicircle, which, as we have just seen, gives no contribution to the
integral. We can now integrate using the residue theorem.

We note that the integrand in \ref{free_num_int} has a simple pole at
$\w=\xi(\bp)-i \eta \ sign(\xi(\bp))$. Thus, if $\xi(\bp)>0$, then
the pole is in the lower complex plane, and the integral vanishes, and
if $\xi(\bp)>0$, then the pole is in the upper complex plane, with
residue 1. Using the residue theorem, and summarizing these results we
finally get
\begin{equation}
n(\bp)=\Theta(-\xi(\bp)),
\end{equation}
which we recognize as the zero temperature fermi distribution.

\subsection{Interacting Green functions}

According to Dyson's equation, the green function for an interacting
system has the form
\begin{equation}
G(w,\bq)={1\over \w -\xi(\bp)- \Sigma(\w,\bp)},
\label{GFDyson}
\end{equation}
where $\Sigma(\w,\bp)$ is an arbitrarily complicated function summarizing
all the interactions in the system, which is known as self energy.

The prescription to find $\Sigma$
is quite straightforward, and it consists of adding all amputated
connected feynman diagrams for the system. The fact that, in general,
the number of such diagrams is infinite, makes this task virtually impossible.
Nonetheless, eq. \ref{GFDyson} is very powerful, since it allows one to
include the effect of
infinite subsets of the total number of diagrams in the system, by
only having to explicitly calculate a few representative ones.

An alternative standard approach, leads to the exact result (NOTE:
Abrikosov measures energy from $\mu$, here we measure from 0, which
is more standard.)
\begin{equation}
G(\w,\bp)=\int_0^\infty{d\w'\left[{A(\w,\bp)\over \w-\w'+i
      \eta}+{B(\w,\bp)\over \w+\w'-i \eta}\right]},
\label{leh}
\end{equation}
Where $A$ and $B$ are, again, arbitrary complicated functions, though
they are known to be finite.

To understand $A$ and $B$ more closely, we need to introduce the
following well known identity:
\begin{equation}
\lim_{\nu \to 0} {1 \over x \pm i \nu}
=\mathcal{P}{1\over x}\mp i \pi  \delta(x),
\label{form}
\end{equation}
where $\mathcal{P}$ is a Cauchy principal value, which represents the
contribution due to a discontinuity in a Riemann sheet (branch cut),
and the delta function represents the contribution due to the pole.

Applying \ref{form} to \ref{leh} we get
\begin{eqnarray}
\label{imleh}
&&Re\ G(\w,\bp)=\mathcal{P}\int_0^\infty{d\w'\left[{A(\w,\bp)\over \w-\w'+i
      \eta}+{B(\w,\bp)\over \w+\w'-i \eta}\right]} \\
&&Im\ G(\w,\bp)=\left\{ \begin{array}{r c c} -\pi A(\w,\bp)&if&\w>0 \\
      \pi B(-\w,\bp)&if&\w<0 \end{array}
\right.
\end{eqnarray}

Finally, eq. \ref{free_num_int} represents a fundamental property of
green functions, and it can be generalized to interacting systems
simply substituting $G^0$ with $G$. Applying it to eq. \ref{leh}, and
performing the $\w$ integral first, we get
\begin{equation}
n(\bp)=\int_0^\infty{d\w' B(\w,\bp)}=\int_{-\infty}^{0}{d\w{-1\over \pi}
  Im\ G(\w,\bp).
}
\end{equation}
Introducing the function $\rho(\w,\bp)=-2 Im G(\w,\bp)$, generally
called spectral function, the above equation can be written as
\begin{equation}
n(\bp)=\int{{d\w\over 2 \pi}\rho(\w,\bp)\Theta(-\w)}.
\label{dist}
\end{equation}

An important property of the spectral function is that for all $\bp$,
\begin{equation}
\int{{d\w\over 2 \pi}\rho(\w,\bp)}=1.
\end{equation}
This can be understood as a sum rule in the following sense: if we
wish to calculate the number of holes in the system, we would take
the $\eta \to 0^-$ limit in equation \ref{free_num_int}. The
distribution would then have been $n_{holes}(\bp)=1-n(\bp)=
\int_0^\infty{d\w' A(\w',\bp)}$, so that $ 1=\int_0^\infty{d\w'
 \left[ A(\w',\bp)+B(\w',\bp)\right]}= \int{{d\w\over 2
    \pi}\rho(\w,\bp)}$.

Using eq.\ref{GFDyson}, together with the definition of $\rho$, we
can write
\begin{equation}
\rho(\w,\bp)={-2 Im \Sigma(\w,\bp) \over
  \left[\w-\xi(\bp)-Re\Sigma(\w,\bp)\right]^2+ \left[Im
  \Sigma(\w,\bp)\right]^2}.
\end{equation}
Furthermore, if $\Sigma$ were to be real, or if, equivalently, the
pole of the green function were to be real, for some momentum $\bp$,
then taking the limit $Im \Sigma \to 0$ of \ref{GFDyson}, and using
eq,\ref{form}, we get
\begin{equation}
\rho(\w,\bp)=2 \pi \delta(\w-\xi(\bp)-Re\Sigma(\w,\bp)),
\end{equation}
which can be simplified, using the properties of the delta function,
to
\begin{equation}
\rho(\w,\bp)=2 \pi Z(\bp) \delta(\w-\w_0(\bp)),
\label{rhoZ}
\end{equation}
where Z, known as spectral weight is given by
\begin{equation}
Z(\bp)={1\over\left|1-{\partial \over \partial
    \w}Re\Sigma(\w,\bp)\right|_{w=\w_0(\bp)}},
\label{Z}
\end{equation}
and $\w_0(\bp)$ is the pole of the green function, defined by
\begin{equation}
\w_0(\bp)-\xi(\bp)-\Sigma(\w_0(\bp),\bp)=0.
\end{equation}
 The momentum distribution in this case is thus given by
\begin{equation}
n(\bp)=Z(\bp)\int{\delta(\w-\w_0(\bp))\Theta(-\w)}=Z(\bp)\Theta(-\w_0(\bp)).
\label{zdens}
\end{equation}

\bibliographystyle{unsrt}

\end{document}